\documentclass[twocolumn,showpacs,preprintnumbers,amsmath,amssymb,floatfix,nofootinbib]{revtex4-1}
\usepackage{amsthm}
\usepackage{amsmath}
\usepackage{graphicx}
\usepackage{dcolumn,xcolor}
\usepackage{bm}
\usepackage[normalem]{ulem}

\newcommand{\be}{\begin{equation}}
\newcommand{\ee}{\end{equation}}
\newcommand{\ba}{\begin{eqnarray}}
\newcommand{\ea}{\end{eqnarray}}
\newcommand{\ban}{\begin{eqnarray*}}
\newcommand{\ean}{\end{eqnarray*}}
\newcommand{\non}{\nonumber}
\newcommand{\eq}[1]{(\ref{#1})}
\newcommand{\n}[1]{\label{#1}}
\newcommand{\hh}{\hspace{0.25cm}}
\newcommand{\hhh}{\, ,\hspace{0.5cm}}
\newcommand{\ind}[1]{\mbox{\tiny #1}}

\newcommand{\cu}{{\cal U}}
\newtheorem{definition}{Definition}

\begin{document}
 
\title{Metamorphoses of a photon sphere}
\author{Andrey A. Shoom}
\email{ashoom@mun.ca, ashoom@ualberta.ca}
\affiliation{Department of Mathematics and Statistics, Memorial University, St. John's, Newfoundland and Labrador, A1C 5S7, Canada}

\begin{abstract}
There are circular planar null geodesics at $r=3M$ around a Schwarzschild black hole of mass $M$. These geodesics form a photon sphere. Null geodesics of the Schwarzschild space-time which do not form the photon sphere are either escape to null infinity or get captured by the black hole. Thus, from the dynamical point of view, the photon sphere represents a smooth basin boundary that separates the basins of escape and capture of the dynamical system governing the null geodesics. Here we consider a Schwarzschild black hole distorted by an external, static, and axisymmetric quadrupolar gravitational field defined by a quadrupole moment $q$. We study null geodesics around such a black hole and show that the photon sphere does not survive the distortion. For $q\lesssim-0.017001$ it transforms into a fractal basin boundary that indicates chaotic behavior of the null geodesics. We calculate the box-counting fractal dimension of the basin boundary and the related uncertainty exponent, which depend on the value of the quadrupole moment.
\end{abstract}


\maketitle

\section{INTRODUCTION}

The Schwarzschild black hole space-time has circular photon trajectories at $r=3M$, where $r$ is the areal radial coordinate and $M>0$ is the black hole mass.\footnote{In this paper we use geometrized units $G=c=1$ and the sign conventions adopted in \cite{MTW}.} These trajectories result from the projection of null geodesics confined to the hypersurface $r=3M$ onto a spatial manifold $t=const$, where $t$ is the Schwarzschild time coordinate. The hypersurface $r=3M$ is unstable in the sense that null geodesics generated by null vectors that are not tangent to the hypersurface diverge from it: they either sink behind the black hole horizon at $r=2M$ or escape to null infinity. This hypersurface is directly related to the gravitational lensing and the black hole shadow \cite{Perlick:2004tq,Perlick:2003vg,Whisker:2004gq,Perlick:2007nt,Virbhadra:2008ws,BinNun:2010ty,Perlick:2015vta,FZ}. The Einstein bending angle becomes arbitrarily large for null rays approaching $r=3M$, and the rim of the shadow is formed by marginally trapped photons that go around the black hole many times before they reach the remote observer. This hypersurface, and also its spatial part, was named a {\em photon sphere}. 

Precise definitions of a photon sphere and more generally, in nonspherically symmetric as well as dynamic cases, of a {\em photon surface}, given in the literature, differ from each other according to the emphasized aspect. For example, in \cite{Virbhadra:1999nm} a photon sphere in static and spherically symmetric space-times is defined as a timelike hypersurface $r=r_0$, such that the Einstein bending angle of a light ray approaching $r=r_0$ becomes infinitely large. In \cite{Claudel:2000yi} is given a geometric definition of a photon surface in a general space-time. A photon surface is defined as an immersed, nowhere spacelike hypersurface, such that for its every point and every null vector tangent to the hypersurface at that point, there exists the corresponding null geodesic lying in the hypersurface. It was also shown that in any static and spherically symmetric space-time, subject to suitable energy conditions, a black hole, a naked singularity, or more than a certain amount of matter must be surrounded by a photon sphere. In \cite{Foertsch:2003ze} a timelike photon 2-surface is defined in a similar way, as a submanifold, not necessarily as a hypersurface. For example, a timelike photon 2-surface in the Schwarzschild black hole space-time is defined by $r=3M$ and $\theta=\pi/2$, where $\theta$ is the polar angle. In \cite{Cederbaum:2014gva}, a photon sphere for an asymptotically flat, static space-time was defined as a timelike embedded hypersurface for which the lapse function is constant. Using this definition, uniqueness theorems for different types of space-times were proven in \cite{Cederbaum:2014gva,Cederbaum:2015aha,Yazadjiev:2015hda,Cederbaum:2015fra,Yazadjiev:2015mta,Yazadjiev:2015jza,Rogatko:2016mho,Tomikawa:2016dqz,Tomikawa:2017vun}. In \cite{Gibbons:2016isj}, a photon surface is defined as a totally geodesic submanifold of the optical manifold in a static space-time. Photon orbits of such a submanifold are unstable. However, in some space-times, such photon orbits are stable and they lie in the so-called antiphoton spheres \cite{Gibbons:2016isj}. In this paper we shall use this definition. Note that in the case of a Kerr black hole there are different types of ``photon surfaces" that depend on the value of azimuthal angular momentum \cite{Teo} and thus, according to the definitions above, they are not photon surfaces. Finally, we would like to mention that a more general concept of a transversely trapping surface was introduced in \cite{Shiromizu:2017ego,Yoshino:2017gqv}. According to the definition, the transversely trapping surface is a timelike, static, or stationary hypersurface, such that photons emitted tangentially to it either propagate along the surface or fall into its interior, which is defined by a two-dimensional and orientable spacelike cross section of the transversely trapping surface. 

Photon surfaces are widely studied. In \cite{Hod:2017xkz} was wound a generic upper bound on black hole photon spheres in static, spherically symmetric, asymptotically flat space-times; $r\leq 3M$, where $r$ is the areal radius of a null circular geodesic and $M$ is the total ADM mass of the space-time. In \cite{Hod:2011aa} it was proven that in the case of hairy black holes, in static, spherically symmetric, and asymptotically flat space-times, that the hair extends beyond the space-time photon sphere, and it was conjectured that the region above the photon sphere contains $\gtrsim 50\%$ of the total hair mass. The upper bound on the black hole photon spheres was extended to higher-dimensional Einstein and Einstein-Gauss-Bonnet space-times in \cite{Gallo:2015bda}. In the eikonal approximation, the quasinormal modes of a black hole in a static, spherically symmetric, and asymptotically flat space-time can be expressed through the frequency and the instability time scale of an unstable photon circular orbit, which is an intersection of the space-time photon sphere and the black hole equatorial plane \cite{Cardoso:2008bp,Stefanov:2010xz}. Note however, that as it was demonstrated in \cite{Konoplya:2017wot}, the expected relation between the frequency and the instability time scale of an unstable photon circular orbit around a black hole and quasinormal modes is violated in the Lovelock theory of gravity. It was also stated that such a relation can exist for any stationary, spherically symmetric, and asymptotically flat black hole if its perturbations are limited to test fields that are minimally coupled to gravity. The nature of a relation between quasinormal modes and null particle orbits was analyzed and criticized in \cite{Khanna:2016yow}. 

Unstable null circular orbits are also closely related to the ringdown waves from a vibrating compact object \cite{Press:1971wr,Goebel,Ferrari:1984zz,Berti:2009kk} and from the binary black hole coalescences observed in the gravitational waves emissions detected by aLIGO so far \cite{Abbott:2016blz,Abbott:2016nmj,TheLIGOScientific:2016qqj,TheLIGOScientific:2016pea,Abbott:2017vtc}. A precise observation of the late-time ringdown signal from a compact binary coalescence should be done to rule out ``exotic" compact objects as alternatives to black holes \cite{Cardoso:2016rao}. In \cite{Decanini:2010fz}, for static and spherically-symmetric black hole space-times, a precise connection was established between a photon sphere and the properties of surface waves associated with Regge poles of the scattering matrix that are propagating close to it. Photon spheres that coincide with an extremal horizon of static, spherically-symmetric black hole space-times were studied in \cite{Khoo:2016xqv}. Photon and antiphoton spheres in static, spherically symmetric-solutions of supergravity theories, a Horndeski theory, and a theory of quintessence were studied in \cite{Cvetic:2016bxi}. Gravitational lensing by naked singularities and the formation of photon spheres around them was studied in \cite{Virbhadra:1998dy,Virbhadra:2002ju,Virbhadra:2007kw}. 

In this paper, we consider a Schwarzschild black hole distorted by the external, static, and axisymmetric gravitational field. The black hole space-time is a member of the Weyl class of solutions (see, e.g., \cite{Synge,Chandrasekhar}).  The distortion field can be induced by massive objects that are not explicitly included in the Weyl solution. 
Thus, this solution is vacuum and describes the space-time in the vicinity of the distorted black hole horizon. Accordingly, this solution represents a {\em local black hole} \cite{Geroch:1982bv}. Here we consider a quadrupole distortion. This type of distortion allows for equatorial null geodesics that were studied in \cite{Shoom:2015slu}. Moreover, many celestial objects, such as stars, star clusters, and galaxies, are nearly symmetric with respect to their equatorial plane. For such a space-time, the escape dynamics of massive and massless test particles and its chaotic nature were analyzed in a different context in \cite{deMoura:1999wf}. Here we consider null geodesics propagating outside the equatorial plane. Our goal is to analyze whether null geodesics form a photon surface around the distorted Schwarzschild black hole. 

The issue of the existence of a photon surface in the presence of static, first-order metric perturbations of the Schwarzschild space-time, retaining asymptotic flatness, was studied in \cite{Yoshino:2016kgi}. It was conjectured that if an asymptotically flat, vacuum space-time possesses a static photon surface, then the space-time is the Schwarzschild one. It was also argued that in the presence of a fine-tuned matter perturbation, a photon surface may exist. Here we explore this issue further and consider an exact, vacuum, nonasymptotically flat solution representing a distorted Schwarzschild black hole.     

This paper is organized as follows. In the next section, following the works \cite{Claudel:2000yi} and \cite{Gibbons:2016isj}, we give the definition of a photon surface. In Sec. III, we present the metric of a distorted Schwarzschild black hole, briefly discuss its main properties, and consider a quadrupole distortion. In Sec. IV, we derive the dynamical system governing null ray trajectories of the distorted Schwarzschild black hole space-time and consider the corresponding effective potential. In Sec. V, we study small oscillations of null ray trajectories around an equatorial null circular orbit. In Sec. VI, we integrate the dynamical system numerically, present its basins of attraction, and study the basin boundary. The conclusion contains summary and discussion the derived results. 

\section{Photon surface}

As we already mentioned in the Introduction, a geometric definition of a photon surface in a general space-time was given in \cite{Claudel:2000yi}. Here we give this definition for a 4D space-time. 
\begin{definition}
A 3D photon surface is an immersed, nowhere-spacelike hypersurface $\Sigma_{\ind{Ph}}$ of a 4D space-time manifold ${\cal M}$ such that, for every point $p\in\Sigma_{\ind{Ph}}$ and every null vector ${\bm k}\in T_p\Sigma_{\ind{Ph}}$,  there exists a null geodesic $\gamma$ such that, ${\bm k}$ is tangent to $\gamma$ at $p$ and $\gamma\subset\Sigma_{\ind{Ph}}$.                         
\end{definition}
In other words, we can pick up any point on a 3D photon surface and a null vector tangent to the surface at that point. Then this vector generates a null geodesic that lies in the photon surface. For example, any null hypersurface is a 3D photon surface, a single-sheeted hyperboloid in 4D Minkowski space-time is a 3D photon surface, the $r=3M$ hypersurface of the Schwarzschild black-hole space-time is a photon surface. 

In the case of a static space-time, that we consider here, if a 3D photon surface does not depend on a time coordinate, its projection onto a spatial manifold is time-independent and there is an associated 2D photon surface in the spatial manifold ${\cal S}:\,t=const$ of the space-time metric
\be
ds^2=-N^2dt^2+g_{ij}dx^idx^j\,,
\ee
where the lapse function $N$ and spatial metric $g_{ij}$ are independent of $t$. Such a 2D photon surface $S_{\ind{Ph}}$ is the result of a spatial projection of the 3D photon surface $\Sigma_{\ind{Ph}}$ onto the spatial manifold ${\cal S}$. The null ray trajectories are spatial geodesics of the optical metric
\be
ds_{\ind{opt}}^2=N^{-2}g_{ij}dx^idx^j\,.
\ee 
They can be constructed by using Fermat's principle \cite{Gibbons:2016isj}. Accordingly, one can define a 2D photon surface as follows:

\begin{figure}[htb]
\begin{center}
\hspace{0cm}
\includegraphics[width=4.5cm]{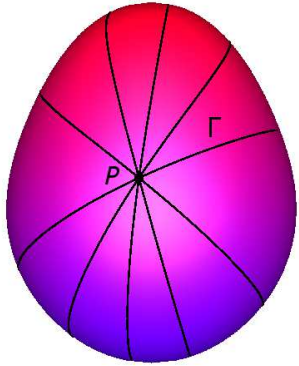}
\caption{2D Photon Surface $S_{\ind{Ph}}$. Through every point $P\in S_{\ind{Ph}}$ and in every direction runs a null ray trajectory $\Gamma\subset S_{\ind{Ph}}$.}\label{fig1}
\end{center}
\end{figure}

\begin{definition}
A 2D photon surface $S_{\ind{Ph}}$ is a totally geodesic hypersurface of the spatial manifold ${\cal S}$, which is a spatial projection of the associated 3D photon surface $\Sigma_{\ind{Ph}}$, such that projection of a null vector ${\bm k}\in T\Sigma_{\ind{Ph}}$ onto $S_{\ind{Ph}}$ is the spacelike vector ${\bm v}\in TS_{\ind{Ph}}$, which generates the null ray trajectory $\Gamma\subset S_{\ind{Ph}}$, which is the projection of the null geodesic $\gamma\subset\Sigma_{\ind{Ph}}$ generated by ${\bm k}$.                         
\end{definition}
Such a 2D photon surface is schematically illustrated in Fig.~\ref{fig1}. A null ray trajectory runs in every direction through every point in a 2D photon surface. 

\section{Distorted Schwarzschild\\ black hole}

\subsection{The metric}

Consider a Schwarzschild black hole in the presence of static and axisymmetric external gravitational field. Focus on the space-time in the exterior vicinity of the black hole horizon. This region of space-time represents a {\em local black hole}. Such a local black hole solution is vacuum, static, and axisymmetric, and it is given by a Weyl solution that admits an asymptotically flat extension that includes the external matter generating the field. Details of this construction and basic characteristics of a local black hole can be found in the paper \cite{Geroch:1982bv}. Here we shall call such a local black hole a distorted Schwarzschild black hole. The metric of a distorted Schwarzschild black hole in prolate spheroidal coordinates $(t,x,y,\phi)$ has the following form: 
\ba
ds^{2}&=&-\left(\frac{x-1}{x+1}\right)e^{2U}dt^{2}+m^{2}(x+1)^{2}(1-y^{2})e^{-2U}d\phi^{2}\non\\
&+&M^{2}(x+1)^{2}e^{2(V-U)}\left(\frac{dx^{2}}{x^{2}-1}+\frac{dy^{2}}{1-y^{2}}\right)\,,\n{II.1a}\\
U&=&\sum_{n\geq0}a_{n}R^{n}P_{n}\,,\n{II.1b}\\
V&=&\sum_{n,k\geq1}\frac{nka_{n}a_{k}}{(n+k)}R^{n+k}(P_{n}P_{k}-P_{n-1}P_{k-1})\,\non\\
&+&\sum_{n\geq1}a_{n}\sum_{l=0}^{n-1}[(-1)^{n-l+1}(x+y)-x+y]R^{l}P_{l}\,,\n{II.1c}\\
P_{n}&\equiv&P_{n}(xy/R)\hhh R=\sqrt{x^{2}+y^{2}-1}\,.\non
\ea
The metric \eq{II.1a} was studied in many works (see, e.g., \cite{ER,Dor,Chandrasekhar,Hoenselaers:1979mk,Quevedo:1989rfm,Manko,Frolov:2007xi}) and the explicit form of the metric function $V$ was given in \cite{BDM}. Here, $P_n$ are the Legendre polynomials of the first kind. The metric functions $U$ and $V$ represent a static and axisymmetric gravitational distortion field defined by the {\em interior Weyl multipole moments} $a_{n}$. In what follows, we shall simply call them multipole moments. The distortion fields $U$ and $V$ defined by the multipole moments are regular and smooth at the black hole horizon, which is located at $x=1$.\footnote{The {\em exterior Weyl multipole moments} describe distortions of the source \cite{BDM,Breton:1998sr}. They are given in terms of the Legendre polynomials of the second kind (see, e.g., \cite{ER,Dor,Hoenselaers:1979mk,Quevedo:1989rfm,Manko,Breton:1998sr}). According to the Schwarzschild black hole uniqueness theorem \cite{Israel:1967wq}, the Schwarzschild black hole is the only static, asymptotically flat, vacuum black hole with a regular horizon. Thus, such distortions make the black hole horizon singular (see, e.g., \cite{Dor,Manko}).} Regular coordinate neighborhoods are defined by the following coordinate ranges: $t\in(-\infty, \infty)$, $x\in(1, \infty)$, $y\in(-1, 1)$, and $\phi\in(0, 2\pi)$. By the construction, this metric diverges at the spatial infinity, $x\to\infty$. 

To have the horizon free of conical singularities at the symmetry axis $y=\pm1$, the multipole moments have to satisfy the following condition:
\be\n{II.7}
\sum_{n\geq0}a_{2n+1}=0\,.
\ee
This condition is sometimes called the black hole equilibrium condition \cite{Chandrasekhar}.

Assuming that the distortion field is generated by some material sources that satisfy the strong energy condition, then we necessarily have (see, e.g., \cite{Geroch:1982bv,Fairhurst:2000xh}), 
\be\n{II.8a}
U\leq0\,,
\ee
which implies 
\be\n{II.8b}
u_{0}=\sum_{n\geq0}a_{2n}\leq0\,.
\ee
 
Using the coordinate transformations 
\be\n{II.9}
x=\frac{r}{M}-1\hhh y=\cos\theta\,,
\ee
and removing the distortion by making all the multipole moments $a_{n}$ vanish we derive the Schwarzschild metric,
\ba\n{II.10}
ds^2&=&-\left(1-\frac{2M}{r}\right)dt^2+\left(1-\frac{2M}{r}\right)^{-1}dr^2+r^{2}d\omega^{2}\,,\non\\
d\omega^{2}&=&d\theta^2+\sin^2\theta d\phi^2\,.
\ea

In what follows, it is convenient to present the metric \eq{II.1a} in a dimensionless form
\ba\n{II.5}
dS^{2}&=&-\left(\frac{x-1}{x+1}\right)e^{2\cu}d\tau^{2}+(x+1)^{2}(1-y^{2})e^{-2\cu}d\phi^{2}\non\\
&+&(x+1)^{2}e^{2(V-\cu)}\left(\frac{dx^{2}}{x^{2}-1}+\frac{dy^{2}}{1-y^{2}}\right)\,,
\ea
where
\ba\n{II.6}
dS^{2}&=&\Omega^{-2}ds^{2}\hhh\Omega^{2}=M^{2}e^{-2u_{0}}\,,\non\\
\tau&=&\frac{t}{M}e^{2u_{0}}\hhh \cu=U-u_{0}\,.
\ea

\subsection{Quadrupole distortion}

The first term in the expansion \eq{II.1b} of the distortion field $U$ is the monopole and in our case it represents a uniform background distortion defined by a monopole moment $a_{0}$. The next term is the dipole defined by a dipole moment $a_{1}$, which according to the black hole equilibrium condition \eq{II.7} is related to the higher-order multipole moments, e.g. $a_1+a_3=0$.  The next term is the quadrupole, which is defined by a quadrupole moment $a_{2}$. It defines the most dominant, nontrivial distortion field that allows for equatorial null geodesics \cite{Shoom:2015slu}, and it also captures the structure of many celestial objects that are nearly symmetric with respect to their equatorial plane. Here we shall consider only the quadrupole distortion and neglect the higher-order multipole moments. According to the expressions \eq{II.1b} and \eq{II.1c}, the quadrupole distortion fields $\cu$ and $V$ read
\ba
\cu&=&\frac{q}{2}\left[3x^2y^2-x^2-y^2-1\right]\,,\n{III.1a}\\\
V&=&\frac{q}{4}(1-y^2)\left[q(x^2-1)(x^2-9x^2y^2+y^2-1)-8x\right]\,,\non\\
\n{III.1b}
\ea
where $q=a_{2}$ is the quadrupole moment. 

To justify the quadrupole approximation, the distortion field should only slightly modify the space-time geometry in the vicinity of the black hole's horizon. This allows us to estimate the magnitude of $q$. To do it, we consider the Kretschmann scalar curvature invariant ${\cal K}\equiv R_{\alpha\beta\gamma\delta}R^{\alpha\beta\gamma\delta}$ calculated on a static black hole horizon,
\be\n{III.2}
{\cal K}|_{\ind{Horizon}}=12K^2\,,
\ee
where $K$ is the Gaussian curvature of the horizon two-dimensional spacelike surface (for details see, e.g., \cite{Frolov:2007xi,Abdolrahimi:2009db}). For the metric \eq{II.5} with the quadrupole distortion fields \eq{III.1a}--\eq{III.1b}, this expression reads
\be\n{III.3}
{\cal K}|_{\ind{Horizon}}=12(1+2q)^2\exp(4q)\,.
\ee
For a Schwarzschild black hole this expression reduces to 12. Thus, for a small quadrupole distortion one should have 
\be\n{III.4}
\left|(1+2q)^2\exp(4q)-1\right|\ll1\,,
\ee
which implies that $|q|\ll1$.

In accordance with the study of the equatorial null geodesics around a distorted Schwarzschild black hole \cite{Shoom:2015slu}, where it was shown that null circular orbits in the black hole equatorial plane exist only if $q\ge q_{\ind{min}}$, where  
\be\n{III.5}
q_{\ind{min}}=\frac{1-2\cos(\pi/9)}{24\cos^2(\pi/9)+20\cos(\pi/9)+2}\simeq -0.020944533\,,
\ee
in what follows, we shall take $q_{\ind{min}}\leq q\ll1$. 

To get more insight into an external quadrupole distortion, let us consider Newtonian gravity.\footnote{Note that the potential $U$ defines a relativistic gravitational field. Its Newtonian limit is given by $\lim_{c^2\to \infty}c^2U(x, y, c^2
)$, where c is the speed of light, which has to be explicitly included into the potential (see, e.g., \cite{Quevedo:1989rfm,Ehlers}).} The Newtonian interior quadrupole moment $q_{\ind{N}}$ of two equal pointlike masses $\mu$ located on the $z-$axis at the distance $d$ from the coordinate origin, and an infinitesimally thin homogeneous ring of the mass $m$ and radius $r$ located in the plane $z=0$ and centred at the origin, is
\be\n{III.6}
q_{\ind{N}}=\frac{m}{2r^3}-\frac{2\mu}{d^3}\,.
\ee  
Hence, if the quadrupole moment due to the masses is not less than that of the ring, $q_{\ind{N}}\leq0$, otherwise, $q_{\ind{N}}>0$. Thus, in analogy with Newtonian gravity, we shall take positive and negative values of $q$. 

\section{Dynamical system\\ for null ray trajectories}

Due to the quadrupole distortion, null geodesics in the vicinity of a distorted Schwarzschild black hole differ from the null geodesics around an isolated Schwarzschild black hole. To analyze the effect of the distortion on the null geodesics, we construct the dynamical system for null ray trajectories. 

Null geodesic equations read
\be\n{IV.1}
\dot{k}^{\alpha}+\Gamma^{\alpha}_{\beta\gamma}k^{\beta}k^{\gamma}=0\,,
\ee
where the 4-velocity vector $k^{\alpha}=(\dot{\tau},\dot{x},\dot{y},\dot{\phi})$ is null,
\be\n{IV.2}
k^{\alpha}k_{\alpha}=0\,,
\ee
and the overdot stands for the derivative with respect to an affine parameter $\lambda$. For the space-time \eq{II.5} the Christoffel symbols $\Gamma^{\alpha}_{\beta\gamma}$ are presented in the Appendix. The null condition \eq{IV.2} takes the following form:
\ba\n{IV.3}
&&-\left(\frac{x-1}{x+1}\right)e^{2\cu}\dot{\tau}^{2}+(x+1)^{2}(1-y^{2})e^{-2\cu}\dot{\phi}^{2}\non\\
&&+(x+1)^{2}e^{2(V-\cu)}\left(\frac{\dot{x}^{2}}{x^{2}-1}+\frac{\dot{y}^{2}}{1-y^{2}}\right)=0\,.
\ea
The space-time \eq{II.5} has Killing vectors $\xi^{\alpha}_{(\tau)}=\delta^{\alpha}_{\tau}$ and $\xi^{\alpha}_{(\phi)}=\delta^{\alpha}_{\phi}$. Accordingly, the following quantities conserved along a null geodesic:
\ba
{\cal E}&\equiv&-\xi^{\alpha}_{(\tau)}k_{\alpha}=\left(\frac{x-1}{x+1}\right)e^{2\cu}\,\dot{\tau}\,,\n{IV.4a}\\
{\cal L}&\equiv&\xi^{\alpha}_{(\phi)}k_{\alpha}=(x+1)^{2}(1-y^{2})e^{-2\cu}\dot{\phi}\,.\n{IV.4b}
\ea

\begin{figure}[htb]
\begin{center}
\hspace{0cm}
\includegraphics[width=7.0cm]{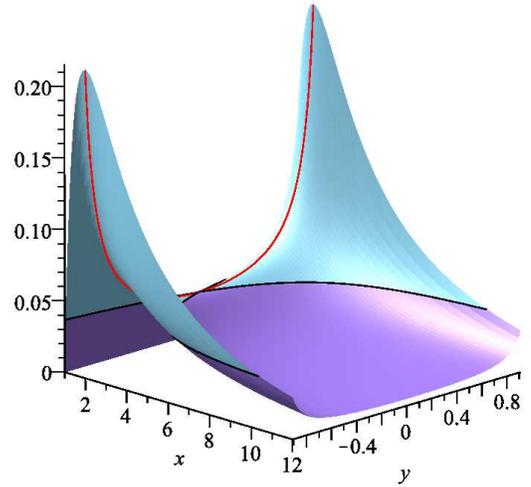}
\caption{The effective potential for a Schwarzschild black hole. Black curves define the local saddle point $(x=2,y=0,U_{\ind{eff}}=1/27)$ level of the effective potential. The red curve represents local extrema $U_{\ind{eff},x}=0$ of the effective potential and defines null ray trajectories at $x=2$ of the 2D photon sphere, which are its large circles.}\label{fig2}
\end{center}
\end{figure}

\begin{figure}[htb]
\begin{center}
\hspace{0cm}
\includegraphics[width=7.0cm]{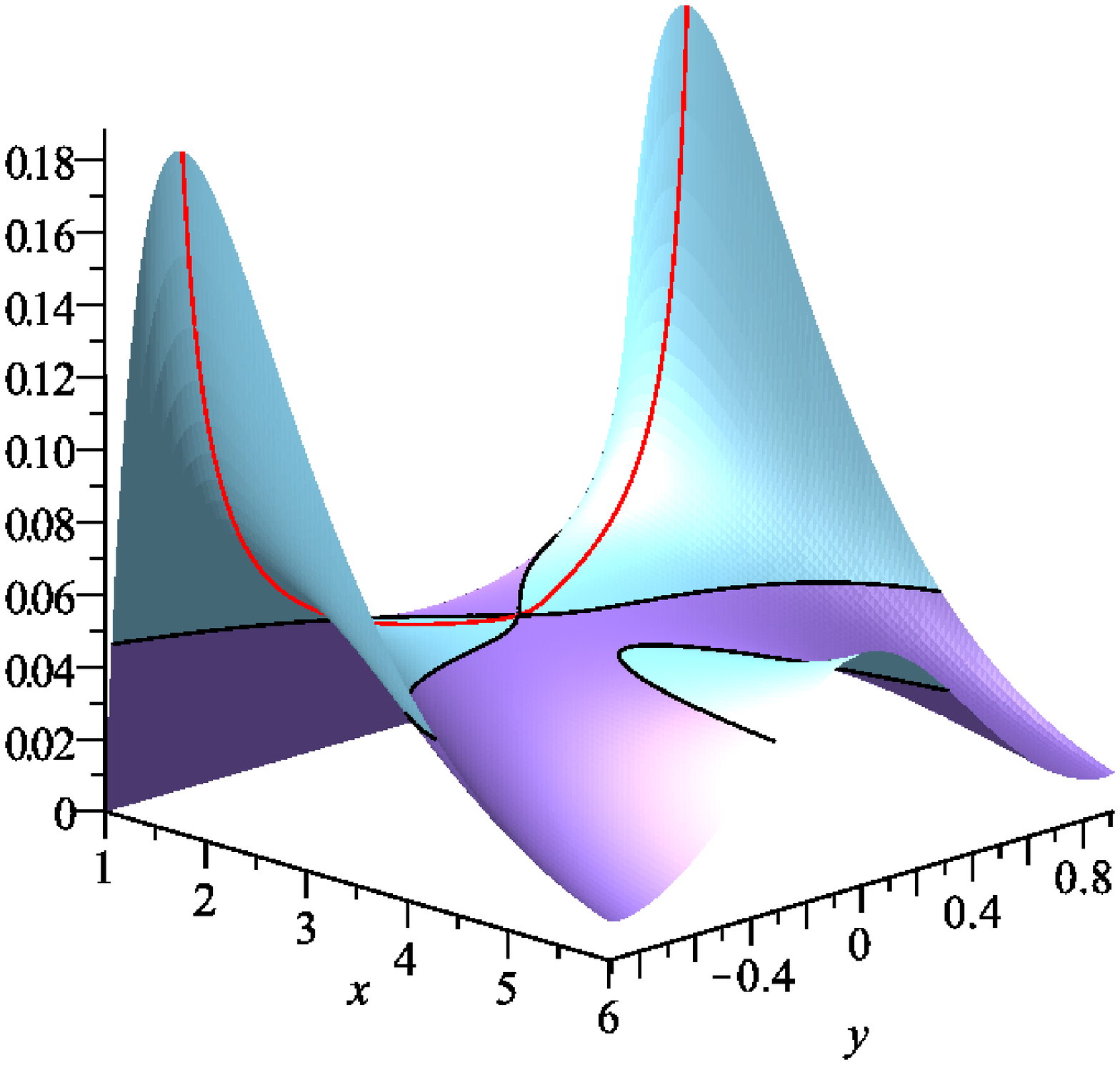}
\caption{The effective potential for a distorted Schwarzschild black hole: $q=-0.02$. Black curves define the local saddle point $(x\simeq2.596,y=0,U_{\ind{eff}}\simeq0.047)$ level of the effective potential. The red curve represents local extrema $U_{\ind{eff},x}=0$ of the effective potential.}\label{fig3}
\end{center}
\end{figure}

\begin{figure}[htb]
\begin{center}
\hspace{0cm}
\includegraphics[width=7.0cm]{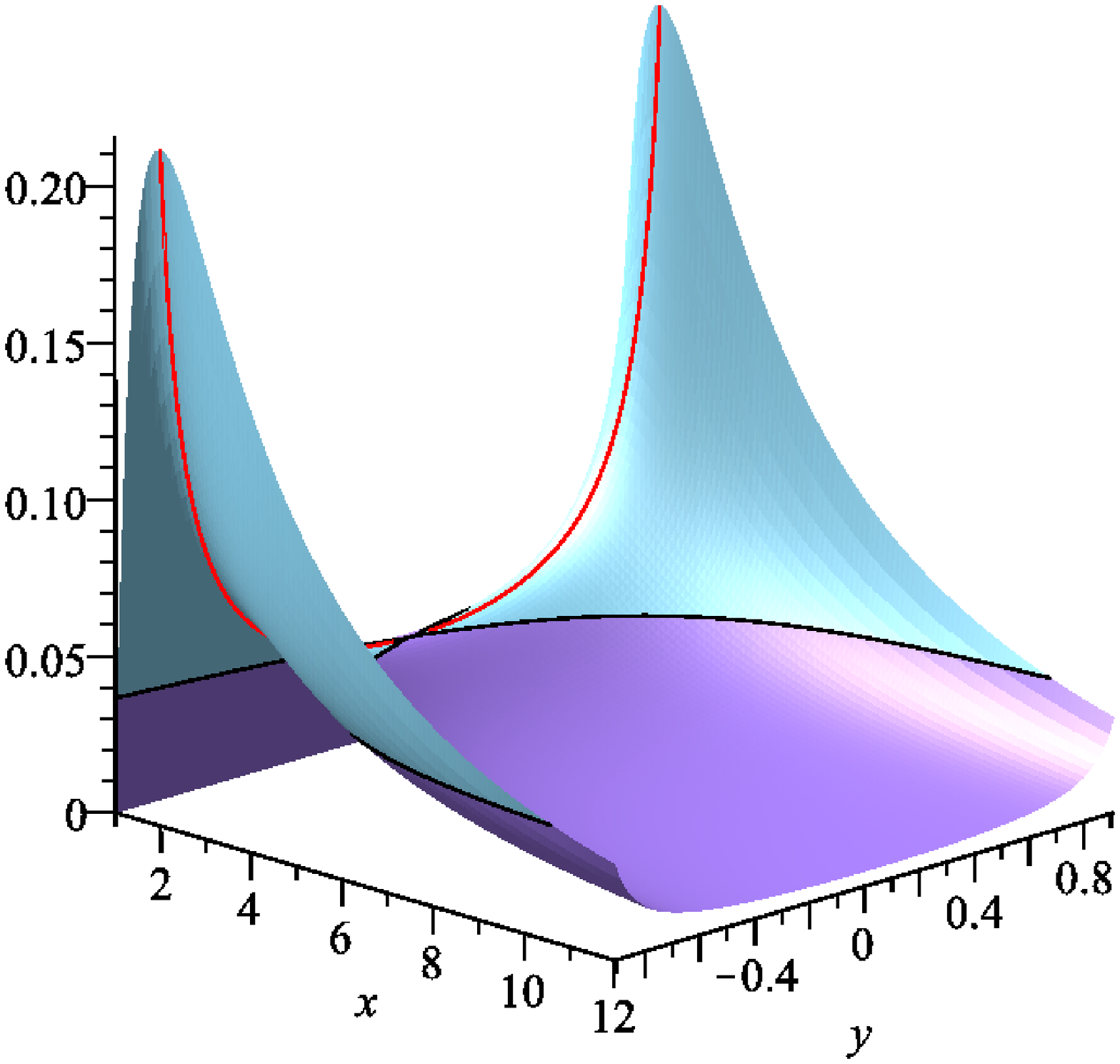}
\caption{The effective potential for a distorted Schwarzschild black hole: $q=0.0001$. Black curves define the local saddle point $(x\simeq1.999,y=0,U_{\ind{eff}}\simeq0.037)$ level of the effective potential. The red curve represents local extrema $U_{\ind{eff},x}=0$ of the effective potential.}\label{fig4}
\end{center}
\end{figure}

In what follows, we shall consider nonradial null geodesics, i.e., $\dot{\phi}\ne0$. According to the expression \eq{IV.4b}, $\phi(\lambda)$ is a monotonic function. As a result, we can use it as an evolution parameter along null ray trajectories. Therefore, we shall consider $\phi$ taking its values in the covering space (see, e.g., \cite{Hawking}), i.e., we shall take $\phi\in[0,\infty)$. Thus, we can divide the expression \eq{IV.3} by $\dot{\phi}^2$ and using \eq{IV.4a} and \eq{IV.4b} present it in the following form:
\be\n{IV.5}
b^{-2}=\frac{(x-1)e^{2(V+2\cu)}}{(x+1)^3(1-y^2)^2}\left[\frac{\left(\frac{dx}{d\phi}\right)^2}{x^{2}-1}+\frac{\left(\frac{dy}{d\phi}\right)^2}{1-y^{2}}\right]+U_{\ind{eff}}\,,
\ee
where $b\equiv{\cal L}/{\cal E}$ is the impact parameter and
\be\n{IV.6}
U_{\ind{eff}}=\frac{(x-1)e^{4\cu}}{(x+1)^3(1-y^2)}\,
\ee
is the effective potential. The effective potential for a Schwarzschild black hole is presented in Fig.~\ref{fig2}, and the effective potential for a distorted Schwarzschild black hole is presented in Figs.~\ref{fig3} and \ref{fig4}.


Taking $\phi$ as an evolution parameter and using the expressions \eq{IV.4a}--\eq{IV.4b}, we construct from the null geodesic equations \eq{IV.1} the dynamical system for null ray trajectories,
\ba
\frac{d^2x}{d\phi^2}&=&-b^{-2}e^{-8\cu}\frac{(x+1)^6(1-y^2)^2}{(x-1)^2}\Gamma^{x}_{\tau\tau}-\tilde{\Gamma}^{x}_{xx}\left(\frac{dx}{d\phi}\right)^2\non\\
&-&2\tilde{\Gamma}^{x}_{xy}\frac{dx}{d\phi}\frac{dy}{d\phi}-\Gamma^{x}_{yy}\left(\frac{dy}{d\phi}\right)^2-\Gamma^{x}_{\phi\phi}\,,\n{IV.7a}\\
\frac{d^2y}{d\phi^2}&=&-b^{-2}e^{-8\cu}\frac{(x+1)^6(1-y^2)^2}{(x-1)^2}\Gamma^{y}_{\tau\tau}-\Gamma^{y}_{xx}\left(\frac{dx}{d\phi}\right)^2\non\\
&-&2\tilde{\Gamma}^{y}_{xy}\frac{dx}{d\phi}\frac{dy}{d\phi}-\tilde{\Gamma}^{y}_{yy}\left(\frac{dy}{d\phi}\right)^2-\Gamma^{y}_{\phi\phi}\,.\n{IV.7b}
\ea
The expression \eq{IV.5} is a constraint along these trajectories. 

In the case of a Schwarzschild black hole, there is a trajectory at $x=2$ that corresponds to the local extrema $U_{\ind{eff},x}=0$ of the effective potential (see the red curve in Fig.~\ref{fig2}). This trajectory defines the 2D photon sphere. A natural question to ask is whether such local extrema of the effective potential of a distorted Schwarzschild black hole (see the red curves in Figs.~\ref{fig3} and \ref{fig4}) also define trajectories of the dynamical system \eq{IV.7a}--\eq{IV.7b}. A positive answer to this question would imply the existence of a photon surface around a distorted Schwarzschild black hole. The curve corresponding to the local extrema is given by
\be\n{IV.8}
2-x-2qx(x^2-1)(1-3y^2)=0\,.
\ee
To find out whether it represents a trajectory of the dynamical system, we shall use the following approach. Let we have a 2D dynamical system in an $(x,y)-$plane
\be\n{IV.9}
\frac{d^2x}{dt^2}=F_x(x,y,v_x,v_y)\hhh \frac{d^2y}{dt^2}=F_y(x,y,v_x,v_y)\,.
\ee
Here $t$ is an evolution parameter (time), $F_x$ and $F_y$ are force components, and
\be\n{IV.10}
\frac{dx}{dt}=v_x\hhh \frac{dy}{dt}=v_y\,.
\ee
are velocity components. Let a curve ${\cal C}$,
\be\n{IV.11} 
{\cal C}:\,\,\,f(x,y)=0\,,
\ee
defines a trial trajectory of the dynamical system \eq{IV.9}. Then, the velocity vector, which is tangent to this curve, satisfies the relation
\be\n{IV.12}
f_{,x}v_x+f_{,y}v_y=0\,.
\ee
Here and in what follows, $(\ldots)_{,x}$ stands for a partial derivative of $(\ldots)$ with respect to $x$. Accordingly, the velocity components can be presented as follows:
\be\n{IV.13}
v_x=af_{,y}\hhh v_y=-af_{,x}\,,
\ee  
where $a=a(x,y)$ is a function to be defined. From the expressions \eq{IV.9}, \eq{IV.10}, and \eq{IV.13} it follows:
\ba
\frac{d^2x}{dt^2}&=&a(af_{,xy}f_{,y}-af_{,yy}f_{,x}+a_{,x}f_{,y}^2-a_{,y}f_{,x}f_{,y})\stackrel{\mathrm{\cal C}}{=}F_x\,,\non\\
\n{IV.14a}\\
\frac{d^2y}{dt^2}&=&a(af_{,xy}f_{,x}-af_{,xx}f_{,y}+a_{,y}f_{,x}^2-a_{,x}f_{,x}f_{,y})\stackrel{\mathrm{\cal C}}{=}F_y\,.\non\\
\n{IV.14b}
\ea 
Here $\stackrel{\mathrm{\cal C}}{=}$ means that the equality holds on the curve ${\cal C}$. To isolate the unknown function $a$, we can multiply \eq{IV.14a} and \eq{IV.14b} by $f_{,x}$ and $f_{,y}$, respectively, and add them to derive
\be\n{IV.15}
a^2(f_{,xx}f_{,y}^2-2f_{,xy}f_{,x}f_{,y}+f_{,yy}f_{,x}^2)+f_{,x}F_x+f_{,y}F_y\stackrel{\mathrm{\cal C}}{=}0\,.
\ee
This expression can also be derived from the expression of a centripetal force acting on a unit mass,
\be\n{IV.16}
F_{c}=\frac{\vec{v}^{\,2}}{r_{c}}\,,
\ee
where $F_{c}$ is the force, which is orthogonal to the velocity $\vec{v}$, and $r_{c}$ is the curvature radius of the mass trajectory. In our case, this expression takes the following form:
\be\n{IV.17}
(\vec{F},\vec{n})=\kappa a^2|\nabla f|^2\,,
\ee
where on the left-hand side we have the projection of the force $\vec{F}$ on a unit vector 
\be\n{IV.18}
\vec{n}=\frac{\nabla f}{|\nabla f|}=\frac{(f_{,x}, f_{,y})}{(f_{,x}^2+f_{,y}^2)^{1/2}}
\ee
orthogonal to the curve ${\cal C}$, and on the right-hand side we have a product of the curvature of the curve 
\be\n{IV.19}
\kappa=-\frac{f_{,xx}f_{,y}^2-2f_{,xy}f_{,x}f_{,y}+f_{,yy}f_{,x}^2}{(f_{,x}^2+f_{,y}^2)^{3/2}} 
\ee
and the squared velocity $v^2=v_x^2+v_y^2$ is expressed by using \eq{IV.13}.

Given a curve \eq{IV.11}, one can solve Eq. \eq{IV.15} for $a$ and check whether the expressions \eq{IV.14a} and \eq{IV.14b} hold on ${\cal C}$. If they do, then the curve represents a true trajectory of the dynamical system \eq{IV.9}. 
 
Applying this method to the dynamical system \eq{IV.7a}--\eq{IV.7b} and the curve \eq{IV.8}, one can see that for $q=0$ the corresponding curve $x=2$ is indeed a true trajectory, while for $q\ne0$ the curve \eq{IV.8} is not a true trajectory. Note that even for very small values of $y$, such that $|y|\ll1$, \eq{IV.14a} is violated in the second order in $|y|$, while \eq{IV.14b} is violated in the first order in $|y|$. 

Are there other candidates for a photon surface around a distorted Schwarzschild black hole? 

\section{Small oscillations}

In \cite{Shoom:2015slu} it was found that there are equatorial null orbits around a Schwarzschild black hole distorted by an external quadrupolar gravitational field. Let us now consider small oscillations about an equatorial null circular orbit defined by $x=x_o$ and $y=0$, where $x_o\in(1,x_{\ind{max}}]$ is the smallest positive root of equation \eq{IV.8} with $y=0$,
\be\n{V.0}
2-x_o-2qx_o(x_o^2-1)=0\,,
\ee
and 
\be\n{V.10}
x_{\ind{max}}=1+2\cos(\pi/9)\approx2.87938524\,,
\ee
is the smallest positive root corresponding to $q_{\ind{min}}$. Let $x_o^\alpha(\lambda)=(\tau_o(\lambda),x_o,0,\phi_o(\lambda))$ be the corresponding null geodesic, $k^\alpha_o=(\dot{\tau}_o,0,0,\dot{\phi}_o)$ its 4-velocity null vector, and 
\be\n{V.1}
x^\alpha=x_o^\alpha+\delta x^\alpha\hhh k^\alpha=k_o^\alpha+\delta\dot{x}^\alpha
\ee
be the neighboring null geodesic and its 4-velocity null vector. Here $\delta x^\alpha$ is a deviation from $x_o^\alpha$ and $\delta\dot{x}^\alpha$ is the corresponding 4-velocity deviation. In what follows, we shall consider very small deviations, $|\delta x^\alpha|\ll 1$, $|\delta\dot{x}^\alpha|\ll1$ and use a linear approximation. Substituting \eq{V.1} into the geodesic equation \eq{IV.1} and the null condition \eq{IV.2}, and expanding in powers of $\delta x^\alpha$ and $\delta\dot{x}^\alpha$, we derive in the first-order approximation
\ba
&&\delta\ddot{x}^\alpha+2\Gamma^\alpha_{\beta\gamma}k_o^\beta\delta\dot{x}^\gamma+\Gamma^\alpha_{\beta\gamma,\delta}\delta x^\delta k_o^\beta k_o^\gamma\approx0\,,\n{V.2a}\\
&&2g_{\alpha\beta}k_o^\alpha\delta\dot{x}^\beta+g_{\alpha\beta,\gamma}\delta x^\gamma k_o^\alpha k_o^\beta\approx0\,,\n{V.2b}
\ea
where the Christoffel symbols and their derivatives are evaluated on the null circular orbit. Using the expressions \eq{IV.4a} and \eq{IV.4b}, the 4-velocity null vector can be presented as follows:
\be\n{V.3}
k_o^\alpha=\left(-{\cal E}_og_{\tau\tau}^{-1},0,0,{\cal L}_og_{\phi\phi}^{-1}\right)\,,
\ee
where the metric tensor components are evaluated on the circular orbit. At this stage, as we already did before, we consider the $\phi_o$ coordinate as an evolution parameter. Using \eq{V.3}, we have
\be\n{V.4}
\delta\dot{x}^\alpha={\cal L}_og_{\phi\phi}^{-1}\frac{d\,\delta x^\alpha}{d\phi_o}\hhh \delta\ddot{x}^\alpha={\cal L}_o^2g_{\phi\phi}^{-2}\frac{d^2\delta x^\alpha}{d\phi_o^2}\,.
\ee
Substituting \eq{V.3} and \eq{V.4} into \eq{V.2a} and \eq{V.2b}, we derive equations for the deviation $\delta x^\alpha$,
\ba
&&\frac{d^2\delta \tau}{d\phi_o^2}+2b_o\Gamma^{\tau}_{\tau a}\frac{d\,\delta x^a}{d\phi_o}=0\,,\n{V.5a}\\
&&\frac{d^2\delta x^a}{d\phi_o^2}+2b_o\Gamma^{a}_{\tau\tau}\frac{d\,\delta \tau}{d\phi_o}+2\Gamma^{a}_{\phi\phi}\frac{d\,\delta \phi}{d\phi_o}+\gamma^a_{\;\;b}\delta x^b=0\,,\n{V.5b}\\
&&\frac{d^2\delta \phi}{d\phi_o^2}+2\Gamma^{\phi}_{a\phi}\frac{d\,\delta x^a}{d\phi_o}=0\,,\n{V.5c}\\
&&\hspace{-0.6cm}2b_og_{\tau\tau}\frac{d\,\delta\tau}{d\phi_o}+2g_{\phi\phi}\frac{d\,\delta\phi}{d\phi_o}+(b_o^2g_{\tau\tau,a}+g_{\phi\phi,a})\delta x^a=0\,,\n{V.5d}
\ea
where
\be\n{V.6}
\gamma^a_{\;\;b}\equiv b_o^2\Gamma^{a}_{\tau\tau,b}+\Gamma^{a}_{\phi\phi,b}\,.
\ee
Here, $b_o\equiv{\cal L}_o/{\cal E}_o$ is the impact parameter corresponding to the circular orbit, and the lowercase Latin indices $a$ and $b$ stand for the coordinates $x$ and $y$. Equations \eq{V.5a} and \eq{V.5c} can readily be integrated,
\ba
\frac{d\,\delta\tau}{d\phi_o}&=&b_oV^\phi_0-2b_o\Gamma^\tau_{\tau a}(\delta x^a-\delta x_0^a)\,,\n{V.7a}\\
\frac{d\,\delta\phi}{d\phi_o}&=&V^\phi_0-2\Gamma^\phi_{a\phi}(\delta x^a-\delta x_0^a)\,,\n{V.7b}
\ea
where $\delta x_0^a=(\delta x_0, \delta y_0)$ are the initial deviations and $V^\phi_0$ is the initial value of $d\,\delta\phi/d\phi_o$. Note that the constants of integration ensure that the expression \eq{V.5d} holds. Substituting the expressions \eq{V.7a} and \eq{V.7b} into Eq. \eq{V.5b}, using the expressions for Christoffel symbols given in the Appendix, and simplifying with the use of Eq. \eq{V.0} we derive
\ba
\frac{d^2\delta x}{d\phi_o^2}+\omega_x^2\delta x=0\,,\n{V.8a}\\
\frac{d^2\delta y}{d\phi_o^2}+\omega_y^2\delta y=0\,.\n{V.8b}
\ea

\begin{figure}[htb]
\begin{center}
\hspace{0cm}
\includegraphics[width=7.0cm]{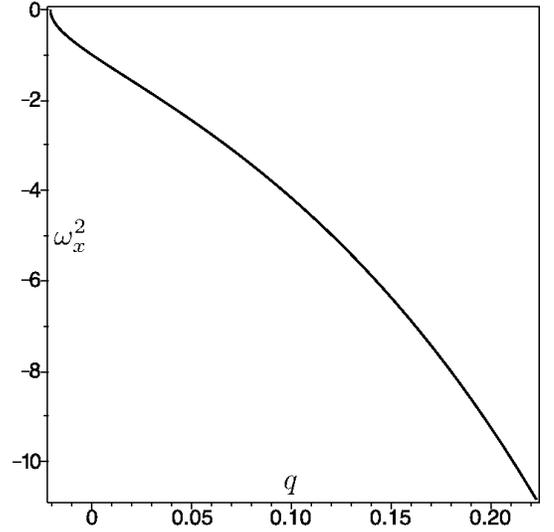}
\caption{The squared frequency $\omega_x^2$ as a function of $q\geq q_{\ind{min}}$.} \label{fig5}
\end{center}
\end{figure}

Here, $\omega_x$ and $\omega_y$ are frequencies of the radial and axial oscillations,
\ba
\omega_x^2&=&\frac{2(x_o^3-3x_o^2+1)}{x_o(x_o^2-1)}e^{-\tfrac{(x_o-2)(x_o^3+14x_o^2-x_o+2)}{8x_o^2(x_o^2-1)}}\,,\n{V.9a}\\
\omega_y^2&=&-\omega_x^2\,.\n{V.9b}
\ea
These expressions allow us to analyze stability of the oscillations. As we can see from Fig.~\ref{fig5}, $\omega_x^2<0$ everywhere except for the marginally stable orbit $x_o=x_{\ind{max}}$, corresponding to $q=q_{\ind{min}}$ [see \eq{III.5}]. Accordingly, $\omega_y^2>0$ everywhere, except for the marginally stable orbit. This implies that for $1<x_o<x_{\ind{max}}$ or $q>q_{\ind{min}}$ small oscillations in the $x-$direction are unstable, while small oscillations in the $y-$direction are stable. According to the expressions \eq{V.7a} and \eq{V.7b}, for a sufficiently small interval in $\phi_o$, both of the oscillations are accompanied by small deviations in $\tau$ and $\phi$. There are no oscillations in the case of the marginally stable orbit.

\section{Numerical analysis} 

\subsection{Numerical setup}

To find out whether there is a photon surface, as it is defined in Sec. II, we shall numerically integrate the dynamical system \eq{IV.7a}--\eq{IV.7b}. To do so, we rewrite it in the first-order form,
\ba
\frac{dx}{d\phi}&=&X\,,\n{VI.1}\\
\frac{dX}{d\phi}&=&-b^{-2}e^{-8\cu}\frac{(x+1)^6(1-y^2)^2}{(x-1)^2}\Gamma^{x}_{\tau\tau}-\tilde{\Gamma}^{x}_{xx}X^2\non\\
&-&2\tilde{\Gamma}^{x}_{xy}XY-\Gamma^{x}_{yy}Y^2-\Gamma^{x}_{\phi\phi}\,,\n{VI.2}\\
\frac{dy}{d\phi}&=&Y\,,\n{VI.3}\\
\frac{dY}{d\phi}&=&-b^{-2}e^{-8\cu}\frac{(x+1)^6(1-y^2)^2}{(x-1)^2}\Gamma^{y}_{\tau\tau}-\Gamma^{y}_{xx}X^2\non\\
&-&2\tilde{\Gamma}^{y}_{xy}XY-\tilde{\Gamma}^{y}_{yy}Y^2-\Gamma^{y}_{\phi\phi}\,.\n{VI.4}
\ea
The constraint \eq{IV.5} reads,
\be\n{VI.5}
\frac{(x-1)e^{2(V+2\cu)}}{(x+1)^3(1-y^2)^2}\left[\frac{X^2}{x^{2}-1}+\frac{Y^2}{1-y^{2}}\right]+U_{\ind{eff}}-b^{-2}=0\,.
\ee 
We also have to specify initial data,
\be\n{VI.6}
x_0=x(\phi_0)\,,\hh X_0=X(\phi_0)\,,\hh y_0=y(\phi_0)\,,\hh Y_0=Y(\phi_0)\,,
\ee
where $\phi_0$ is the initial value of the azimuthal angle. Because the system is axisymmetric, we can always take $\phi_0=0$. 

According to the definition of a 2D photon surface, we can take {\em any} tangent vector ${\bm v}_0$ at the arbitrary point of the surface, then the null ray trajectory generated by ${\bm v}_0$ will lie in it. It means that the trajectory will neither be captured by the black hole, nor will it escape out. However, when we search for a photon surface, we do not know either the point that belongs to it, or the surface orientation that allows us to define a tangent to it vector ${\bm v}_0$. But, because of arbitrary direction of the tangent vector ${\bm v}_0$ at any point of the surface, we can always take it directed in the $\phi-$direction. Accordingly, we consider the following initial data:
\be\n{VI.7}
x_0=x(\phi_0)\,,\hh X_0=0\,,\hh y_0=y(\phi_0)\,,\hh Y_0=0\,,
\ee
that correspond to a turning point of a null ray trajectory in an $(x,y)-$domain. If the generated null ray trajectory gets captured by the black hole or escapes out, then the initial point $(x_0,y_0)$ does not belong to the sought photon sphere. In other words, according to the definition of a photon sphere, it is enough to rule out one particular null ray trajectory generated at the given point by some tangent vector to claim that the point does not belong to the surface.       

Using the initial data \eq{VI.7} we can calculate from the constraint \eq{VI.5} the value of $b$. This constraint also allows us to find the absolute accuracy of the numerical integration. To integrate the dynamical system, we implement the fifth-order quality-controlled Cash-Karp-Runge-Kutta method with the relative accuracy $\varepsilon\lesssim10^{-11}$ (see \cite{PTVF}). We define the absolute error $\delta$ of the numerical integration as the maximal absolute deviation in the constraint \eq{VI.5} among all the points of all generated null trajectories. 

\subsection{Basins of attraction}

To search for a photon surface, we shall scan an entire area in the $(x,y)-$domain in the vicinity of the black hole horizon located at $x=1$. In particular, according to the shape of the effective potential (see Figs.~\ref{fig3} and \ref{fig4}), we consider $x\in(1,3)$. In this region the effective potential is mostly due to the black hole, while for larger values of $x$ the distortion field grows and its effect becomes dominant. Moreover, because we consider small values of the quadrupole moment, a photon surface, if it exists, should not deviate much from the location of a photon sphere at $x=2$ around a Schwarzschild black hole.

\begin{figure}[ht]
\begin{center}
\hspace{0cm}
\includegraphics[width=7.0cm]{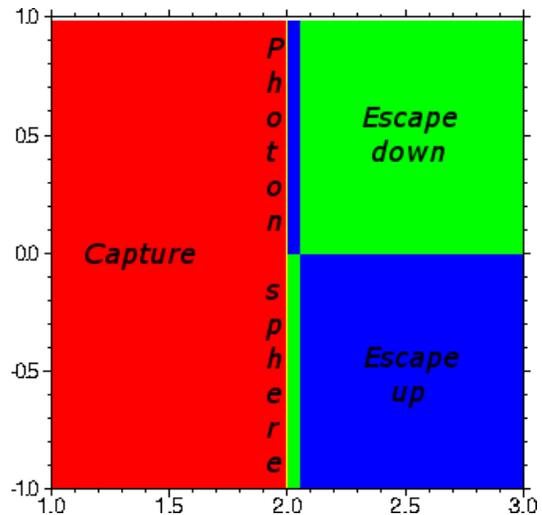}
\caption{Basins of attraction for a Schwarzschild black hole. The relative accuracy is $\varepsilon=10^{-12}$ and the maximal absolute error is $\delta\approx2.03\times10^{-9}$. The horizontal coordinate is $x_0$, the vertical coordinate is $y_0$, and their increment is $\Delta=10^{-2}$. The photon sphere is indicated by yellow vertical line at $x_0=2.0$.} \label{fig6}
\end{center}
\end{figure}
\begin{figure}[ht]
\begin{center}
\hspace{0cm}
\ba
&&\includegraphics[width=7.0cm]{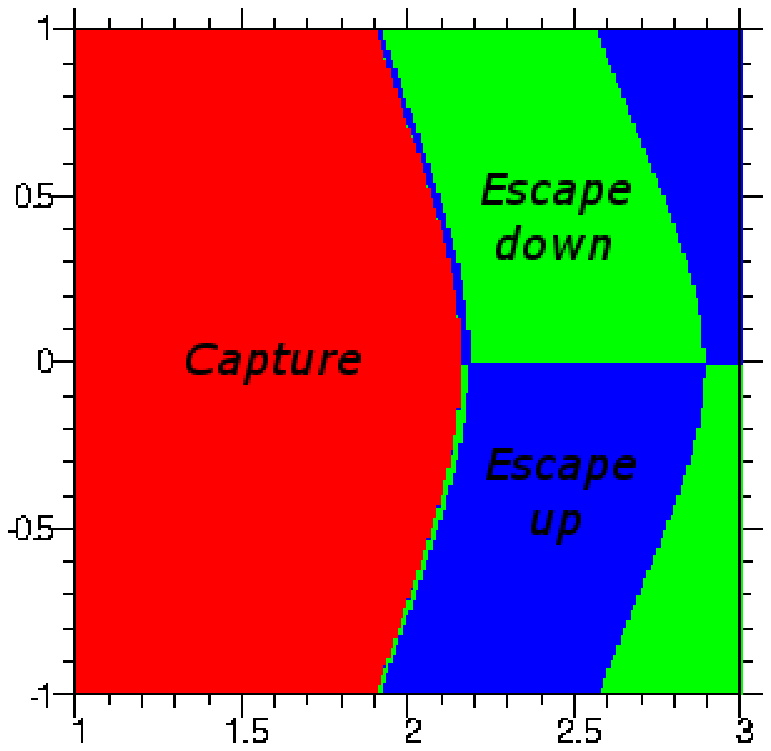}\non\\
&&\hspace{3.5cm}({\bf a})\non\\
&&\includegraphics[width=7.0cm]{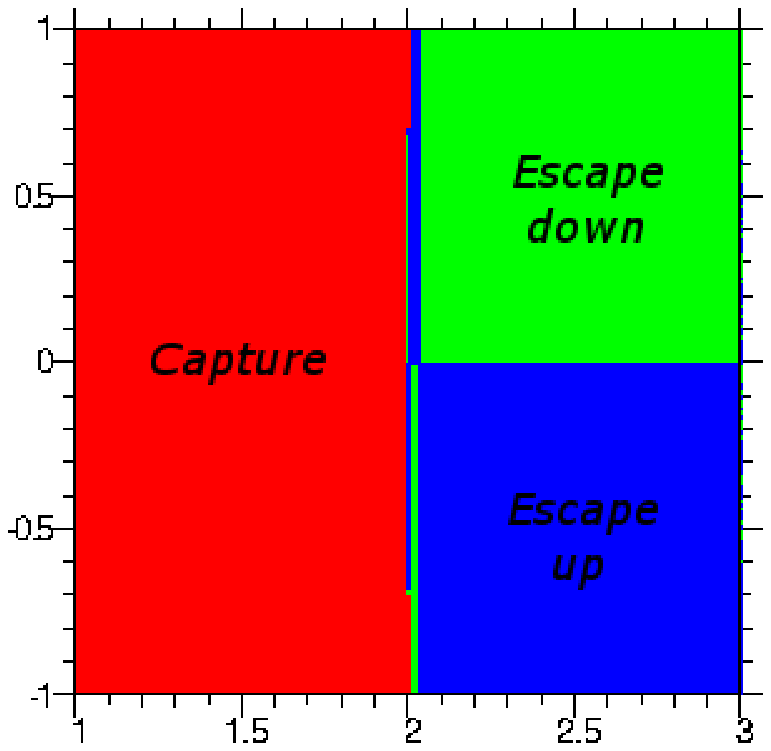}\non\\
&&\hspace{3.5cm}({\bf b})\non
\ea
\caption{Basins of attraction for a distorted Schwarzschild black hole. The relative accuracy is $\varepsilon$ and the maximal absolute error is $\delta$. The horizontal coordinate is $x_0$, the vertical coordinate is $y_0$, and their increment is $\Delta=10^{-2}$.   ({\bf a}): $q=-0.01$, $\varepsilon=10^{-15}$, $\delta\approx5.50\times10^{-12}$. ({\bf b}): $q=0.0001$, $\varepsilon=10^{-14}$, $\delta\approx1.22\times10^{-12}$.}\label{fig7}
\end{center}
\end{figure}
\begin{figure*}[p]
\begin{center}
\ba
&&\hspace{1.1cm}\includegraphics[width=6.3cm]{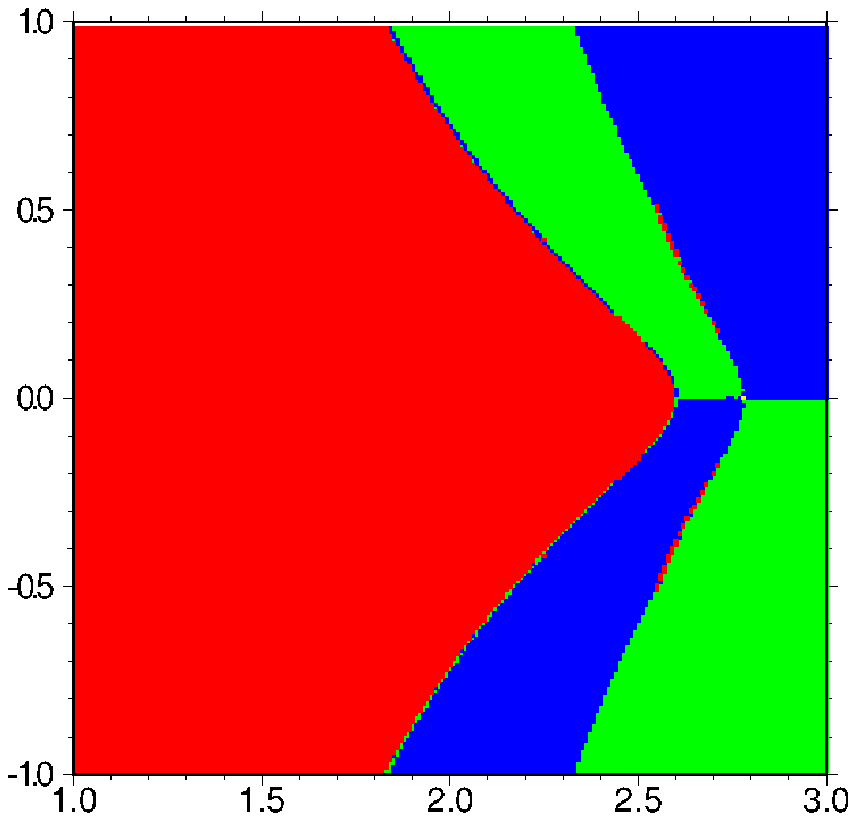}
\hspace{3.0cm}\includegraphics[width=6.65cm]{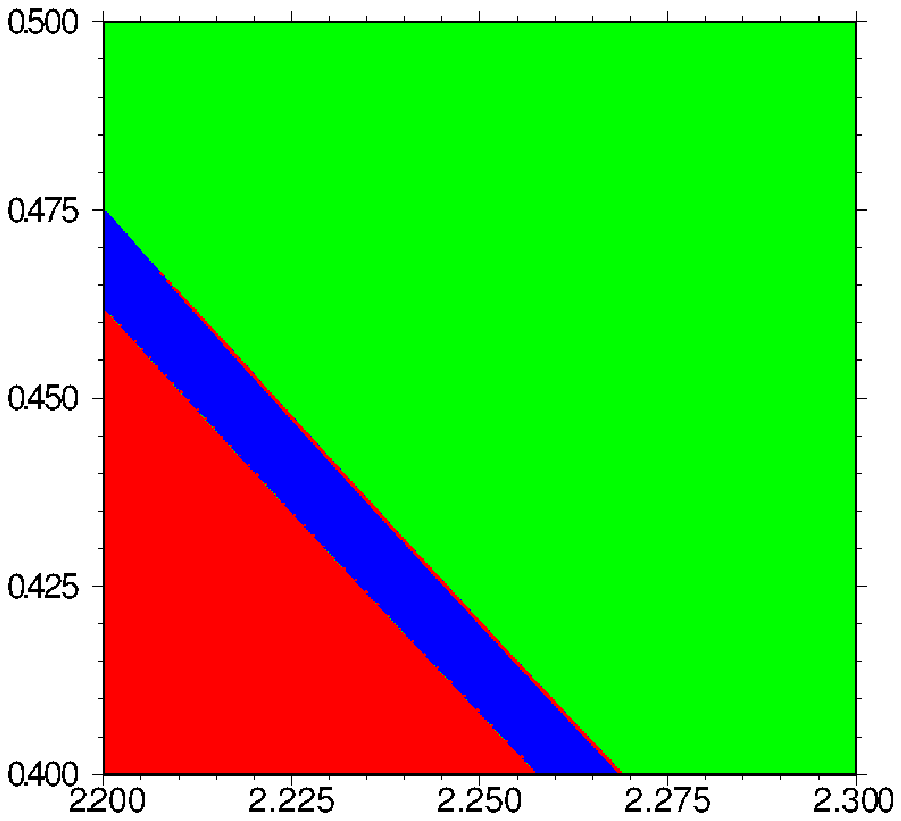}\non\\
&&\hspace{4.2cm}({\bf a})\hspace{9.1cm}({\bf b})\non\\
&&\hspace{0.52cm}\includegraphics[width=7.16cm]{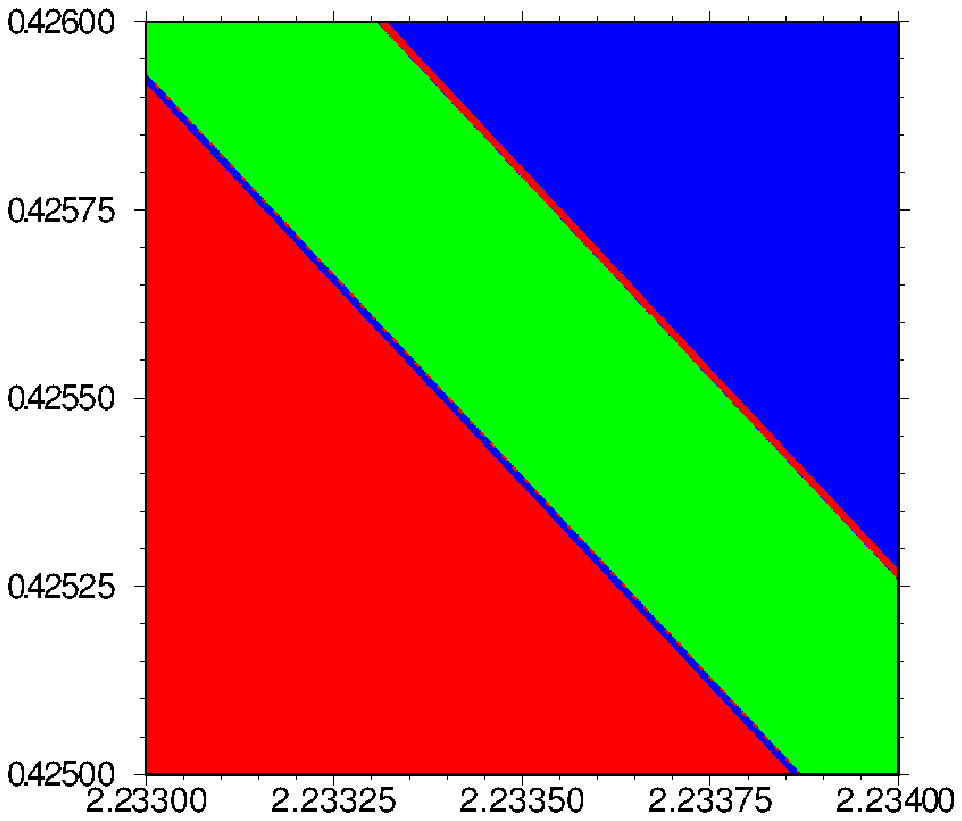}
\hspace{2.2cm}\includegraphics[width=7.5cm]{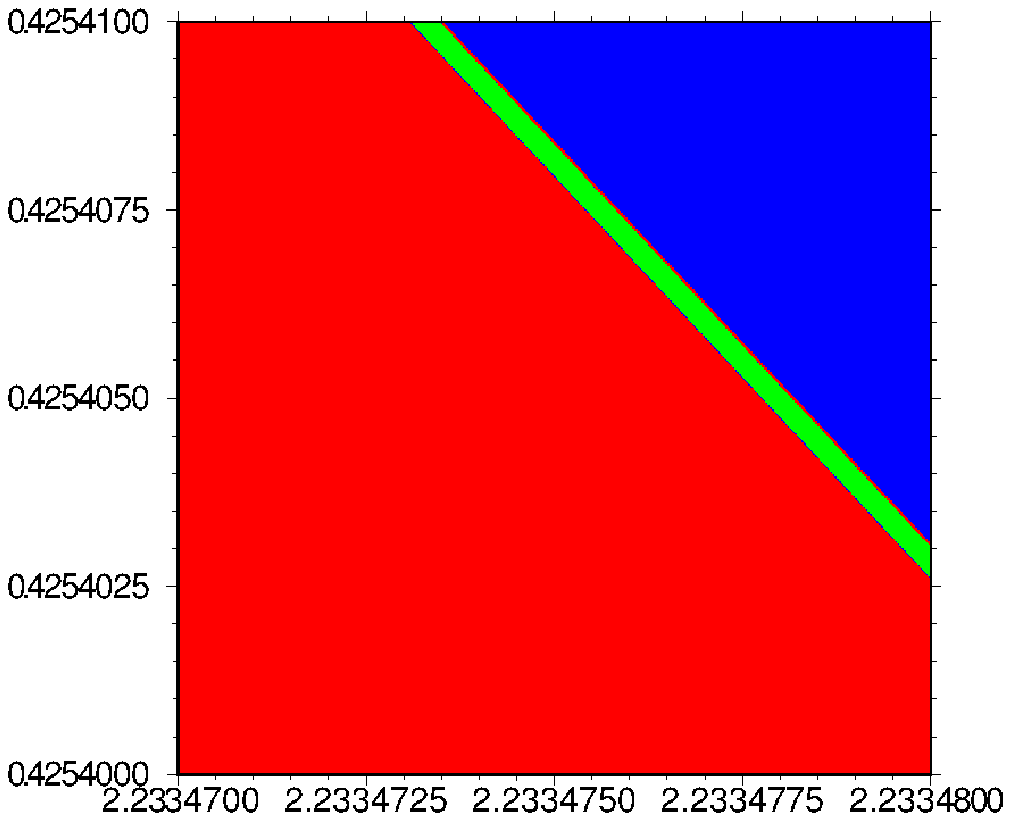}\non\\
&&\hspace{4.2cm}({\bf c})\hspace{9.1cm}({\bf d})\non\\
&&\hspace{0.2cm}\includegraphics[width=7.67cm]{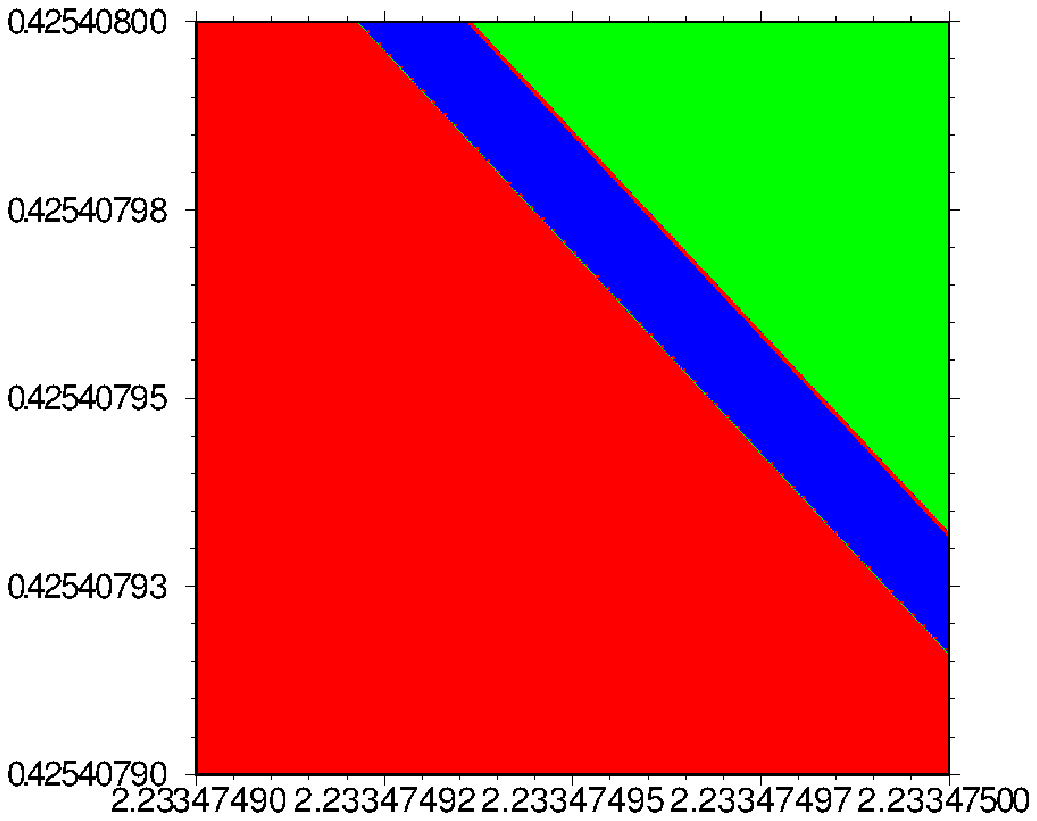}
\hspace{1.7cm}\includegraphics[width=8.0cm]{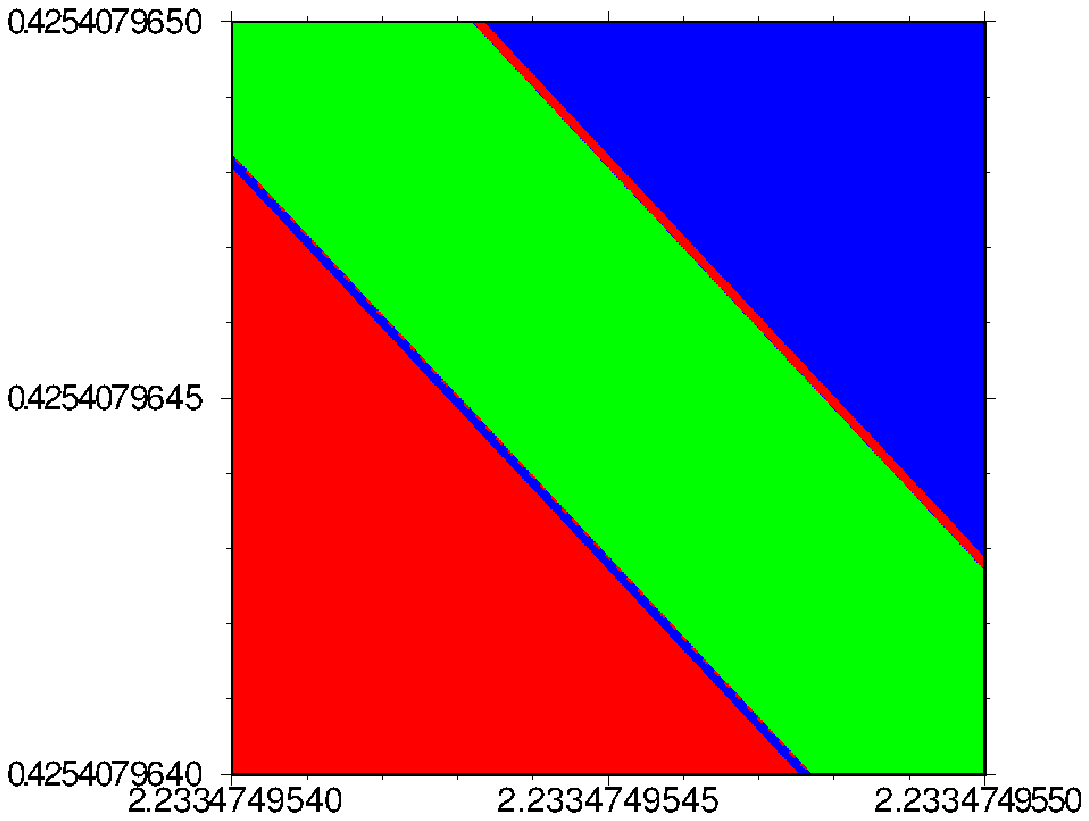}\non\\
&&\hspace{4.2cm}({\bf e})\hspace{9.1cm}({\bf f})\non
\ea
\caption{Basins of attraction for $q=-0.02$. The relative accuracy is $\varepsilon$ and the maximal absolute error is $\delta$. The horizontal coordinate is $x_0$, the vertical coordinate is $y_0$, and their increment is $\Delta$.   ({\bf a}): $\varepsilon=10^{-15}$, $\delta\approx1.20\times10^{-12}$, $\Delta=10^{-2}$. ({\bf b}): $\varepsilon=10^{-13}$, $\delta\approx1.62\times10^{-14}$, $\Delta=10^{-4}$. ({\bf c}): $\varepsilon=10^{-13}$, $\delta\approx1.63\times10^{-14}$, $\Delta=10^{-6}$. ({\bf d}): $\varepsilon=10^{-13}$, $\delta\approx1.64\times10^{-14}$, $\Delta=10^{-8}$. ({\bf e}): $\varepsilon=10^{-13}$, $\delta\approx1.67\times10^{-14}$, $\Delta=10^{-10}$. ({\bf f}): $\varepsilon=10^{-13}$, $\delta\approx1.69\times10^{-14}$, $\Delta=10^{-12}$.}\label{fig8}
\end{center}
\end{figure*}

The scan is done in the following way. We choose a point $(x_0,y_0)$ from the domain 
\be\n{VI.8}
{\cal D}=\{1<x_0<3; -1<y_0<1\}\,
\ee
and using the initial data \eq{VI.7}, we integrate the dynamical system up to sufficiently large values of $\phi:\,50\leq\phi\leq250$. If the result of the integration shows that 
\be\n{VI.9}
x<1.0001\hhh X<0\,,
\ee
then we interpret it that the null ray trajectory got captured by the black hole. Accordingly, we print in ${\cal D}$ a red square of the edge length $\Delta\ll1$, centered at that point and indicating that the trajectory got captured by the black hole. If the result of the integration shows that     
\be\n{VI.10} 
x\geq5\hhh X>0\hhh y<0\,,
\ee 
then we interpret it that the null ray trajectory escaped the black hole in the downward direction and print in ${\cal D}$ a green square of the edge length $\Delta\ll1$ centered at that point. If the result of the integration shows that   
\be\n{VI.11}
x\geq5\hhh X>0\hhh y>0\,,
\ee 
then we interpret it that the null ray trajectory escaped the black hole in the upward direction and print in ${\cal D}$ a blue square of the edge length $\Delta\ll1$ centered at that point. Finally, if the result of the integration shows that none of the above outcomes happened, then we interpret it that the null ray trajectory belongs to a 2D photon surface and print in ${\cal D}$ a yellow square of the edge length $\Delta\ll1$ centered at that point. Then, we move to the next point, by making the increment $\Delta$ in the $x-$ and $y-$directions, and repeat the procedure until we cover all of the domain. 

The conditions \eq{VI.9}--\eq{VI.11} define attractors of the dynamical system \eq{VI.1}--\eq{VI.5}.\footnote{More rigorously, an {\em attractor} is a stable asymptotic final state of a dynamical system.} The set of initial conditions for trajectories of the dynamical system that asymptotically approach its attractor is called the {\em basin of attraction} of that attractor. The method described above allows us to construct basins of attraction. To test the method, we integrate the dynamical system in the absence of distortion ($q=0$) that corresponds to a Schwarzschild black hole. As a result, we construct the basins of attraction shown in Fig.~\ref{fig6}. The boundary separating the basin of capture and the basins of escape is a smooth line $x_0=2.0$. This line represents the photon sphere around the Schwarzschild black hole. If we sufficiently magnify the vicinity of the boundary by decreasing the increment $\Delta$, we can see many stripes representing the regions of upward and downward escapes. These regions alternate and get smaller and smaller in size when one approaches the boundary from the outer side. There is only one region of capture in the inner side of the basin boundary.

Integrating the dynamical system for nonzero values of $q$, we have not observed any indication of a photon surface. There are, however, unstable null circular orbits lying in the equatorial plane of a distorted black hole for $q>q_{\ind{min}}$ and equatorial null finite stable orbits for $q\in(q_{\ind{min}},0)$. This result is in accordance with the analysis of equatorial null geodesics presented in \cite{Shoom:2015slu}. For $q\gtrsim-0.017001$ the basin boundary has a similar structure as in the $q=0$ case. Approaching the basin boundary between the capture and the escape regions, we found many stripes representing the regions of upward and downward escapes located in the outer side of the boundary. These stripes alternate and get smaller and smaller in size when one approaches the boundary from the outer side. Just like in the absence of distortion, there is only one region of capture in its inner side. Basins of attraction for $q=-0.01$ and $q=0.0001$ are shown in Fig.~\ref{fig7}.
 
Finally, for $q\lesssim-0.017001$ the basins of attraction pattern changes. In such cases the basin boundary has a fractal structure: when one approaches the basin boundary from the outer side, there are infinitely many regions corresponding to the capture and both types of escapes. This behavior is illustrated in Fig.~\ref{fig8} for $q=-0.02$.

Such a basin boundary is called a {\em fractal basin boundary}. The fractal nature is a result of the chaotic motion of null ray trajectories originating from such a boundary \cite{MGOY,Ott,GOY1,GOY2}. Thus, as a result of the distortion, the smooth basin boundary (the photon sphere) undergoes metamorphoses that lead to its conversion into the basin boundaries that are fractal \cite{GOY1,GOY2,Ott}. In the next subsection, we calculate the box-counting fractal dimension and the uncertainty exponent that are qualitative characteristics of fractal basin boundaries.

\subsection{The uncertainty exponent and box-counting fractal dimension}  

The fractal nature of a fractal basin boundary leads to the increased sensitivity of final states to the initial conditions error. Namely, a small uncertainty $\epsilon\ll 1$ in the initial conditions can dramatically affect the predictability of the final state. The probability $\rho(\epsilon)$ of making a mistake in the final state determination is calculated as the fraction of the area of the phase space located within $\epsilon$ distance of the basin boundary \cite{MGOY,Ott}. This fraction scales as
\be\n{VI.12}
\rho(\epsilon)\sim\epsilon^\alpha\hhh \alpha=D-D_{\ind{B}}\,,
\ee
where $\alpha$ is the uncertainty exponent, $D$ is the dimension of the phase space (the initial conditions space), and $D_{\ind{B}}\geq D-1$ is the box-counting dimension of the basin boundary,
\be\n{VI.13}
D_{\ind{B}}=\lim_{\epsilon\to0}\frac{\ln N(\epsilon)}{\ln(1/\epsilon)}\,,
\ee
where $N(\epsilon)$ is the number of cubes (squares) of the edge length $\epsilon$ needed to cover the boundary and it is assumed that the limit exists. If a basin boundary is nonfractal, then $D_{\ind{B}}=D-1$ and $\alpha=1$. For a fractal basin boundary $D_{\ind{B}}>D-1$ and $\alpha<1$.  In such a case, we have the final state sensitivity. 

\begin{figure}[htb]
\begin{center}
\hspace{0cm}
\includegraphics[width=7.5cm,height=7.5cm]{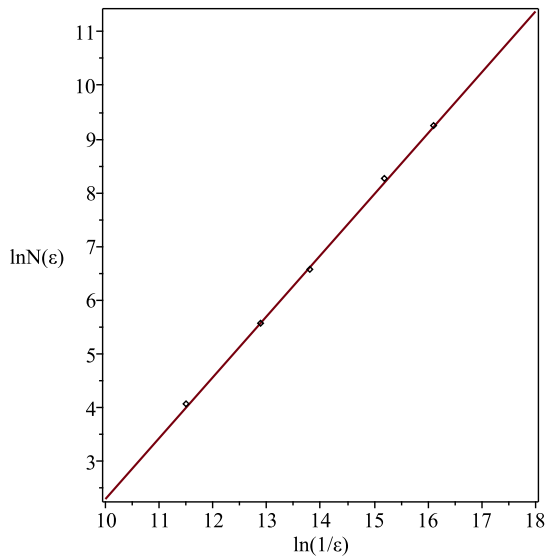}
\caption{The box-counting fractal dimension for $q=-0.02$.} \label{fig9}
\end{center}
\end{figure}

Let us now calculate the box-counting dimension and the uncertainty exponent for the quadrupole distortions of $q=-0.02$. We pick up a small region in the domain ${\cal D}=\{x_0\in[2.23335, 2.23385]; y_0\in[0.42525, 0.42575]\}$ containing the basin boundary. For the given uncertainty $\epsilon=\Delta$, where $\Delta$ is the initial conditions increment, we calculate the number of squares of the edge length $\Delta$ that cover the boundary. These squares are adjacent squares that belong to the basin of capture and the basins of escape. For example, if we pick up a square that belongs to the basin of capture and the next square, at the distance $\epsilon$ from it, that belongs to the basin of upward escape, we say that these points cover the boundary. We compute the number $N(\epsilon)$ of such cases and decrease $\epsilon$ to repeat the procedure.  According to the expression \eq{VI.12}, for a sufficiently small $\epsilon$, a relation between $\ln N(\epsilon)$ and $\ln(1/\epsilon)$ should be linear. Using the linear least squares method, we construct the line representing the relation and calculate its slope that gives us the sought value of $D_{\ind{B}}$ corresponding to $D=2$. Results of such a computation are presented in Fig.~\ref{fig9}. We have
\be
D_{\ind{B}}\simeq1.1343\,,\hh\alpha\simeq0.8657\,.\n{VI.14}
\ee 
The maximal absolute error is $\delta\approx1.73\times10^{-14}$. We have $\alpha<1$. Thus, the basin boundary is fractal. In our computation, we calculated the values of $N(\epsilon)$ for the edge lengths $\epsilon=\{10^{-5},2.5\times10^{-6},10^{-6},2.5\times10^{-7},10^{-7}\}$. 

In general, the box-counting dimension and the uncertainty exponent depend on the chosen region of a basin boundary, however, for a certain class of dynamical systems satisfying required properties, they are proven to be unique \cite{GNOY}. Note that in practice, a dynamical system may not satisfy the required properties to have a unique basin-boundary dimension. In such a case, we may need to calculate $D_{\ind{B}}$ at different regions of the basin boundary. 

\section{Conclusion}

In this paper, we studied null geodesics around a Schwarzschild black hole in the external, static, and axisymmetric quadrupolar gravitational field defined by a quadrupole moment $q$. In particular, we were interested in how such an external field affects the photon sphere around a Schwarzschild black hole. From the dynamical point of view, the photon sphere represents a smooth basin boundary, separating the basin of capture and the basins of escape for the null geodesics of the Schwarzschild space-time. We found that within the numerically accuracy there is no photon surface around such a distorted Schwarzschild black hole. This result implies that the photon sphere around a Schwarzschild black hole is extremely fragile with respect to the external distortion. Despite the fragile nature of the photon sphere, its equatorial cross section, formed by null circular orbits, survives the distortion. Moreover, for $q\in [q_{\ind{min}},0)$, there are finite equatorial stable null orbits outside the black hole. Such orbits were already studied in \cite{Shoom:2015slu}. We also analyzed small oscillations about equatorial null circular orbits, and we found that oscillations in the direction orthogonal to the equatorial plane are stable, while oscillations along the plane in the direction perpendicular to the orbits are unstable.

Our study showed that the quadrupolar distortion of the quadrupole moment $q\lesssim-0.017001$ transforms the basin boundary into a fractal basin boundary. The fractal nature of the basin boundary is illustrated quantitatively by calculating the box-counting dimension of the boundary and the related uncertainty exponent. It indicates a chaotic behavior of null geodesics around the distorted black hole. These quantities and the boundary location depend on the quadrupole moment. As one can see from Fig.~\ref{fig8}, the fractal structure of the boundary is made of points that belong to the capture, downward, and upward escape basins. Such a transformation of a basin boundary is known as {\em basin boundary metamorphoses} \cite{Ott,GOY1,GOY2}.  

The uniqueness theorem for the photon sphere, defined by the constant lapse function, around a Schwarzschild black hole was proven in \cite{Cederbaum:2014gva}, and in a similar way, the uniqueness of a Schwarzschild black hole solution was proven in \cite{Israel:1967wq}. In order to have a photon sphere and a regular black hole horizon, both the proofs require the space-time to be static, vacuum, and asymptotically flat. However, as it is well known, the horizon can be regular if one considers nonvacuum or nonasymptotically flat space-times. In this work, we showed that the situation with a photon sphere is different. Namely, a photon surface ceases to exist, in the sense of the definition given in Sec. II, around a Schwarzschild black hole distorted by the external, quadrupolar gravitational field.\footnote{Note that the corresponding space-time metric \eq{II.1a}--\eq{II.1c} is vacuum but not asymptotically flat.} One may try to explore an arbitrary external, static, and axisymmetric, multipolar gravitational distortion. Our preliminary computations, based on the ``umbilical" definition of a photon surface (see \cite{Claudel:2000yi,Yoshino:2016kgi}), for a hexadecapole field showed that a photon surface does not exist either. A general proof for an arbitrary, external multipolar gravitational distortion remains to be done. 

Finally, we would like to note that the chaotic behavior of test particles is not a rare phenomenon that takes place in strong gravitational fields around black holes. For example, the chaotic motion of massive test particles, in both Newtonian and relativistic core-shell models, represented by a distorted Schwarzschild black hole space-time and its Newtonian limit, was explored in \cite{Vieira:1998hc}. The chaotic behaviour of timelike, null, and spacelike geodesics was discovered in nonhomogeneous vacuum $pp-$gravitational wave space-time \cite{Podolsky:1998ez}. A quantitative and invariant chaos description in some relativistic mechanical systems was given in \cite{Szydlowski:1996uf,Szydlowski:1997jp}. A comparison of symplectic integrator with other nonsymplectic integration schemes for 
nonintegrable relativistic dynamical systems was presented in \cite{Kopacek:2016jdz}. The chaotic motion of timelike particles and light in the Majumdar-Papapetrou multi-black-hole space-times was studied in \cite{Chandra,Contopoulos,Yurtsever:1994yb,Hanan:2006uf,Alonso:2007ts}. And in particular, chaotic scattering of null geodesics and the related analysis of back hole shadows in a binary Majumdar-Papapetrou space-time was studied in \cite{Shipley:2016omi}, and a transition to chaos for stable photon orbits around Reissner-Nordstr\"om diholes was studied with the aid of Poincar\'e sections in \cite{Dolan:2016bxj}. It was demonstrated that gravitational lensing of light by boson stars and Kerr black holes with scalar hair leads to a fractal structure of images of the celestial sphere \cite{Cunha:2015yba,Cunha:2016bjh}. Lensing dynamics of light and shadow of a hairy black hole was studied in \cite{Cunha:2017eoe,Grover:2017mhm}. The story does not end here, and there is much more to be explored in the field of relativistic nonlinear dynamics.

\begin{acknowledgments}

The author is thankful to Volker Perlick and Efthimia Deligianni for criticism and useful comments. This research was supported  by the Natural Sciences and Engineering Research Council of Canada Discovery Grant No. 261429-2013.

\end{acknowledgments}
 
\appendix*

\section{Christoffel symbols}

Here we list Christoffel symbols for the metric \eq{II.5}.
\allowdisplaybreaks
\ba\n{A1}
\Gamma^{x}_{\tau\tau}&=&\frac {( x-1)}{( x+1) ^{3}}\left[1+ ({x}^{2}-1)\,\cu_{,x}\right]e^{4\cu-2V}\,,\non\\
\Gamma^{x}_{xx}&=&-\frac{1}{x^{2}-1}+V_{,x}-\cu_{,x}\,,\non\\
\tilde{\Gamma}^{x}_{xx}&=&\Gamma^{x}_{xx}+2\,\cu_{,x}-\frac{2}{x+1}=\cu_{,x}+V_{,x}-\frac{2x-1}{x^2-1}\,,\non\\
\Gamma^{x}_{xy}&=&V_{,y}-\cu_{,y}\,,\non\\
\tilde{\Gamma}^{x}_{xy}&=&\Gamma^{x}_{xy}+\cu_{,y}+\frac{y}{1-y^2}=V_{,y}+\frac{y}{1-y^2}\,,\non\\
\Gamma^{x}_{yy}&=&-\frac {( x-1)}{(1-y^{2})}\left[1+(x+1)(V_{,x}-\cu_{,x})\right]\,,\non\\
\Gamma^{x}_{\phi\phi}&=&-(x-1)(1-y^{2})\left[1-(x+1)\,\cu_{,x}\right]e^{-2V}\,,\non\\
\Gamma^{y}_{\tau\tau}&=&\frac{( x-1)}{(x+1) ^{3}}(1-{y}^{2})\,\cu_{,y}e^{4\cu-2V}\,,\non\\ 
\Gamma^{y}_{xx}&=&\frac{(1-y^{2})}{(x^{2}-1)}\left(\cu_{,y} -V_{,y}\right)\,,\non\\
\Gamma^{y}_{xy}&=&\frac{1}{x+1}+V_{,x}-\cu_{,x}\,,\non\\
\tilde{\Gamma}^{y}_{xy}&=&\Gamma^{y}_{xy}+\cu_{,x}-\frac{1}{x+1}=V_{,x}\,,\non\\
\Gamma^{y}_{yy}&=&\frac{y}{1-y^{2}}+V_{,y}-\cu_{,y}\,,\non\\
\tilde{\Gamma}^{y}_{yy}&=&\Gamma^{y}_{yy}+2\,\cu_{,y}+\frac{2y}{1-y^{2}}=\cu_{,y}+V_{,y}+\frac{3y}{1-y^{2}}\,,\non\\
\Gamma^{y}_{\phi\phi}&=&(1-y^{2})\left[y+(1-y^{2})\,\cu_{,y}\right]e^{-2V}\,,\non\\
\Gamma^{\tau}_{\tau x}&=&\frac{1}{x^2-1}+\cu_{,x}\,,\non\\
\Gamma^{\tau}_{\tau y}&=&\cu_{,y}\,,\non\\
\Gamma^{\phi}_{x\phi}&=&\frac{1}{x+1}-\cu_{,x}\,,\non\\
\Gamma^{\phi}_{y\phi}&=&-\frac{y}{1-y^2}-\cu_{,y}\,.\non
\ea

\end{document}